\documentclass[fleqn,usenatbib]{mnras}

\usepackage{newtxtext,newtxmath}
\usepackage[T1]{fontenc}
\usepackage[normalem]{ulem} % \sout command
\usepackage[dvipsnames]{xcolor}
\usepackage[flushleft]{threeparttable}
\usepackage{hyperref}
\usepackage{multirow}

\DeclareRobustCommand{\VAN}[3]{#2}
\let\VANthebibliography\thebibliography
\def\thebibliography{\DeclareRobustCommand{\VAN}[3]{##3}\VANthebibliography}

\usepackage{graphicx}	% Including figure files
\usepackage{amsmath}	% Advanced maths commands
\usepackage{tablefootnote}

\usepackage{makecell}

\newcommand {\bb}{\textcolor{black}} 
\usepackage{lipsum}

\title[IM homogeneity scale]{Accessing the homogeneity scale with 21~cm intensity mapping surveys}

\author[Bizarria et al.]{
Bruno B. Bizarria$^{1,2}$\thanks{E-mail: bruno.bizarria@postgrad.manchester.ac.uk}, 
Camila P. Novaes$^{2}$, 
Felipe Avila$^{3}$, 
Rahima Mokeddem$^{2}$, 
Helissa H. da Costa$^{2}$, 
\newauthor
Carlos A. Wuensche$^{2}$,
Gabriel A. S. Silva$^{2}$
\\
$^{1}$Jodrell Bank Centre for Astrophysics, Department of Physics and Astronomy, The University of Manchester, Manchester M13 9PL, UK\\
$^{2}$Instituto Nacional de Pesquisas Espaciais, Av. dos Astronautas 1758, Jardim da Granja, S\~ao Jos\'e dos Campos, SP, Brazil\\
$^{3}$Observatório Nacional, Rua General José Cristino 77, São Cristóvão, 20921-400 Rio de Janeiro, RJ, Brazil
}
\date{Accepted XXX. Received YYY; in original form ZZZ}

\pubyear{\the\year{}}

\begin{document}
\label{firstpage}
\pagerange{\pageref{firstpage}--\pageref{lastpage}}
\maketitle

\begin{abstract}
The homogeneity scale, $R_{\rm H}$, offers a fundamental test of the Cosmological Principle, yet it has not been measured with 21\,cm intensity mapping (IM) surveys. A key limitation for such a measurement is the telescope beam, which artificially smooths the observed signal. We quantify this effect using the two-point correlation function and the correlation dimension, $\mathcal{D}_2(r)$, to model how beam convolution suppresses intrinsic clustering. For any given redshift $z$, we identify a maximum beam width, $\sigma_{\rm max}(z)$, beyond which the homogeneity scale cannot be recovered. This limit defines an inaccessible region in the $\sigma - z$ parameter space for homogeneity measurements.
Applying this framework to several current and upcoming radio telescopes, we assess their ability to probe $R_{\rm H}$.
Our results provide the first quantitative forecast of the instrumental requirements for measuring the cosmic homogeneity scale with 21\,cm IM, and establish a theoretical basis for future observational applications.
\end{abstract}


%%%%%%%%%%%%%%%%%%%%%%%%%%%%%%%%%%%%%%%%%%%%%%%%%%

%%%%%%%%%%%%%%%%% BODY OF PAPER %%%%%%%%%%%%%%%%%%

%%%%%%%%%%%%%%%%%%%%%%%%%%%%%%%%%%%%%%%%%%%%%%%%%%
\section{Introduction}
\label{sec:introduction}
%%%%%%%%%%%%%%%%%%%%%%%%%%%%%%%%%%%%%%%%%%%%%%%%%%

The Cosmological Principle (CP), a central assumption and cornerstone of the standard $\Lambda$CDM cosmological model, states that the Universe is statistically homogeneous and isotropic on sufficiently large scales. Testing the validity of the CP through observations is therefore of great importance, especially given the current challenges faced by the $\Lambda$CDM framework, including the well-known tensions in the measurements of the Hubble constant, $H_0$, and the clustering amplitude, $S_8$ \citep{abdalla2022, diValentino2025}. Unlike isotropy, which has been directly tested in the literature using different observational data \citep{2020/planck-isotropy, 2019/colin, 2019/bengaly, 2018/novaes}, studies of spatial homogeneity are limited to consistency tests. This limitation arises because we only have access to the past light cone, restricted to its intersection with hypersurfaces of constant time \citep{2010/clarkson_maartens, 2011/maartens}. This can be addressed by investigating the existence of a characteristic transition scale to homogeneity, often referred to as the homogeneity scale, beyond which the Universe becomes statistically homogeneous and consistent with the CP. 

The homogeneity scale, $R_{\rm H}$, has already been measured in several galaxy and quasar surveys \citep{Scrimgeour:2012, 2016/laurent, 2021/goncalves, 2018/avila, 2019/avila, franco2025homogeneityscalelocaluniverse}, and has also been proposed as a cosmic ruler for probing cosmology \citep{Ntelis:2018, Xyao:2024}. However, to date, no study has investigated the feasibility of measuring $R_{\rm H}$ using intensity mapping (IM) surveys. In particular, IM of the redshifted 21~cm line emission from neutral hydrogen (rest frequency $\approx 1420$ MHz) is a promising technique for tracing the large-scale structure of the Universe. This method integrates the signal from the 21 cm emission line across radio frequencies, directly related to the redshift of the emitting source, allowing for a tomographic reconstruction of the cosmic volume corresponding to the measured redshift interval. 

Clustering measurements with 21\,cm IM have already been performed through cross-correlation with optical galaxy catalogues, a strategy that mitigates systematics and enhances the signal-to-noise ratio \citep{Pen:2009, Masui:2013, Chang:2010, Switzer:2013, Wolz_gbt:2017, Anderson_cross:2018, Tramonte_parke:2020, Li_parkes:2021, Wolz_gbt:2022, Cunnington_meerkat:2023, Amiri_chime:2023, Amiri_chime:2024, 2025/meerklass_hi}. Given the large number of current and upcoming radio surveys designed to explore this observational window \citep{2012/tianlai, 2014/chime, santos2017meerklass, 2020/ska, 2021/wuensche_BINGO-instrument}, studying this tracer as an alternative probe of the CP is highly relevant. However, extracting cosmological information from 21\,cm IM observations faces major challenges, particularly in foreground subtraction and control of instrumental systematics \cite[see][and references therein]{carucci2024hydrogen}. In this work, we focus on the impact of the angular resolution set by the telescope beam in single-dish radiotelescopes, assessing for the first time the detectability of the homogeneity scale with future 21\,cm surveys. 

%%%%%%%%%%%%%%%%%%%%%%%%%%%%%%%%%%%%%%%%%%%%%%%%%%%%%%%%%%%%%%%%%%%%%%%%%%
\section{Beam effects in configuration space}
\label{sec:beam_effects}
%%%%%%%%%%%%%%%%%%%%%%%%%%%%%%%%%%%%%%%%%%%%%%%%%%%%%%%%%%%%%%%%%%%%%%%%%%%

Homogeneity scale measurements often rely on the use of estimators based on the two-point correlation function (2pCF), $\xi(r)$. Recent work has begun to explore the advantages of the 2pCF for IM surveys, as a configuration-space analysis can circumvent some complexities of Fourier analysis, like ringing and mode-coupling artifacts. \citet{2022/bingo_bao_novaes} presented a forecasting study for Baryon Acoustic Oscillations (BAO) detection from 21\,cm IM simulations using the two-point angular correlation function (2pACF) for the BINGO telescope. \citet{avila2022h} also used HI IM simulations to study the 2pCF, focusing on the BAO signal and the impact of observational effects like beam smoothing and foreground removal. In parallel, \citet{2021/kennedy} investigated the recovery of the BAO scale for a MeerKAT-like survey using the multipoles of the 3D redshift-space correlation function. A significant challenge in such analyses is the proper estimation of the correlated covariance matrix. This can be robustly addressed either through the use of numerous mock simulations, as demonstrated by \citet{2022/bingo_bao_novaes}, or via analytic modelling of the covariance, as implemented by \citet{2021/kennedy}. In the present work, we follow a theoretical approach demonstrating the impact of the radio telescope beam on the 2pCF and on the detectability of the homogeneity scale.

The signal measured by a radio telescope corresponds to the convolution, in configuration space, of the true sky signal with the instrumental beam.
%The signal measured by a radio telescope is a convolution of the true sky signal with the instrument beam in configuration space.
According to the convolution theorem, this real-space convolution translates into a simple multiplicative relationship in Fourier-space \citep{jasche2009digital}.
% \cite[while a convolution in Fourier space translates into a multiplicative relationship in real-space; see][for details]{jasche2009digital}. 
Consequently, the observed power spectrum is given by the intrinsic sky power spectrum multiplied by the squared modulus of the Fourier transform of the beam profile, introducing a scale-dependent damping of the clustering signal \citep{Blake:2019}. This damping primarily suppresses power in the transverse Fourier modes ($k_{\perp}$). While frequency-dependent features can be expected to  affect the parallel modes ($k_{\parallel}$) in the beam profile \citep{Matshawule:2021}, this can be regarded as second order effect in the present study, since the  analysis of the transverse modes contains sufficient information to access the transition to homogeneity. 

We model the instrumental response as a Gaussian shaped beam, characterized by its angular standard deviation \citep{Matshawule:2021, Cunnington_meerkat:2023}
\begin{equation}
\label{eq:beam_width}
    \bb{\sigma(\nu)\,=\, \frac{1.16}{2\sqrt{2\ln 2}} \frac{c }{D_{\rm dish}\nu}},
\end{equation}
which is related to the full width at half-maximum (FWHM) through $\sigma(\nu) = \theta_{\rm FWHM}(\nu)/2\sqrt{2\ln 2}$. Here, $D_{\rm dish}$ is the diameter of telescope dish, $c$ is the speed of light, and $\nu$ is the operating frequency of the instrument.

For cosmological analyses, it is convenient to express the beam width in comoving units. An angular scale $\sigma(\nu)$ at redshift $z$ corresponds to a comoving transverse length $R_{\rm beam} = \chi_c(z)\sigma(\nu)$, where $\chi_c(z)$ is the comoving distance. This quantity sets the characteristic physical scale below which transverse clustering information is smoothed by the instrumental beam.

The power spectrum of the 21\,cm brightness temperature fluctuations, accounting for the damping effect of the beam, is modelled as \citep{Blake:2019, Cunnington_meerkat:2023}
\begin{equation}\label{eq:p_hi}
    P_{\rm HI}^{\rm obs}(k,\mu)\,=\, \mathcal{B}^2(k,\mu)\bar{T}_{\rm HI}^2b_{\rm HI}^2(1+\beta\mu^2)^2 P_m(k),
\end{equation}

\noindent where $P_m(k)$ is the matter power spectrum, $\mu$ is the cosine of the angle between the wave vector $k$ and the line-of-sight, $\bar{T}_{\rm HI}(z)$ the mean brightness temperature, and $b_{\rm HI}$ the HI linear bias. The linear redshift-space distortions (RSD) are incorporated through the Kaiser factor $(1+\beta\mu^2)^2$, with $\beta=f/b_{\rm HI}$ and $f$ the linear growth rate \bb{\citep{kaiser1987clustering}}. The beam damping function $\mathcal{B}(k,\mu)$ is given by 
\begin{equation}
    \mathcal{B}(k,\mu)\,=\,\exp\left[  -\frac{k^2(1-\mu^2)R^2_{\rm beam}}{2}\right].
\end{equation}

The Finger-of-God (FoG) effect \citep{fog:1972} is neglected in this analysis, as it primarily affects small, non-linear scales where the clustering signal is already strongly suppressed by the instrumental beam, and are therefore not expected to impact our analysis. For other cosmological applications, such as using the homogeneity scale $R_{\rm H}$ as a standard ruler, both Kaiser and FoG effects may play an important role and must be taken into account \citep{Ntelis17, Nesseris:2019, shao:2025}.

\begin{figure}
	\includegraphics[width=\linewidth]{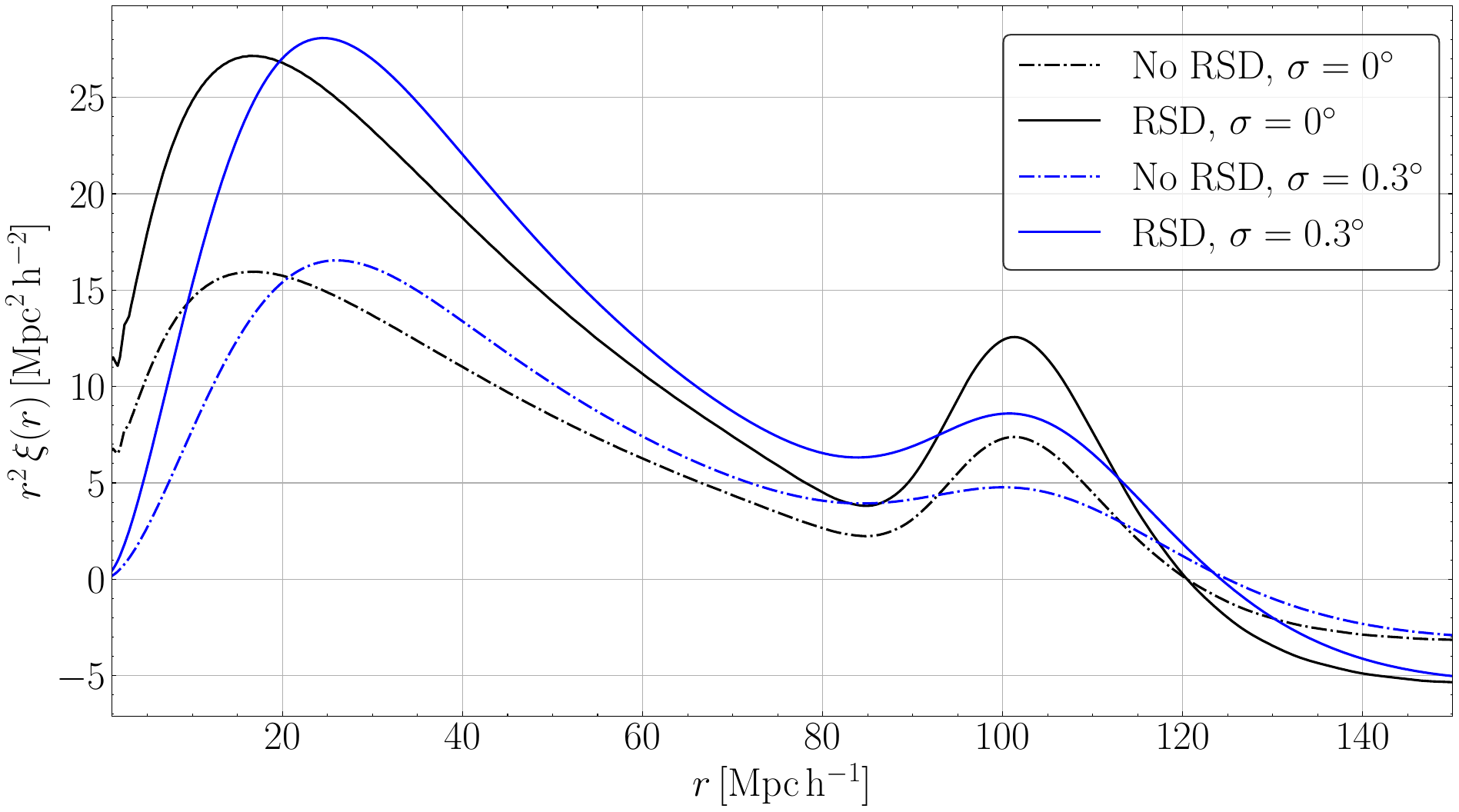}
   \caption{Effect of redshift-space distortions on the 2pCF. We assumed z = 0.8.}
   \label{fig:rsd}
\end{figure}

The straightforward multiplicative relationship between clustering and beam in Fourier space translates into a coupled integral expression in configuration space, where the beam and clustering contributions cannot be factorized. The observed correlation function (monopole moment) is modelled as
\begin{equation}
\begin{split}
\label{eq:integral}
    \xi_{\rm HI}^{\rm obs}(r,\bar{z})\,=&\,\frac{\bar{T}_{\rm HI}^2(\bar{z})b_{\rm HI}^2(\bar{z})}{4\pi^2} \times \\&\int_{-1}^1 \int_{0}^{\infty} d\mu\, dk\, k^2 j_0(k r)\mathcal{B}^2(k,\mu)(1+\beta\mu^2)^2 P_m(k,\bar{z}),
\end{split}
\end{equation}
where $r$ is the separation between pixel pairs, $\bar{z}$ is the mean redshift and $j_0$ is the Bessel function of order zero. Since we focus on the monopole of the 2pCF, linear RSD introduce an enhancement of the clustering amplitude, commonly described by the Kaiser factor, while the instrumental beam primarily modifies the shape of the correlation function by suppressing transverse modes. Figure~\ref{fig:rsd} illustrates the impact of the beam, as well as a comparison between cases with and without the Kaiser effect. Since 21\,cm IM surveys probe the matter distribution in redshift space, all our instrumental forecasts are performed in redshift space unless otherwise stated.

We make use of the semi-numerical code {\tt CLASS}\footnote{\url{http://class-code.net/}} \citep{class_code} to calculate the matter power spectrum, $P_m(k,\bar{z})$. We adopt a fiducial $\Lambda$CDM cosmology consistent with \citet{2018/planck-VI}, with the corresponding parameters listed in Table~\ref{tab:param}, and adopt the {\tt Halofit} model for the non-linear corrections on the power spectrum \citep{Halofit:2003, Takahashi:2012}.
\begin{table}
	\centering
	\caption{Cosmological parameters used in the analysis: Hubble constant $H_0$, cold dark matter density parameter $\Omega_c$, baryon density parameter $\Omega_b$, linear matter fluctuation within $8\, h^{-1}$ Mpc radius $\sigma_8$, and spectral index $n_s$.}
	\label{tab:param}
	\begin{tabular}{lr} 
		\hline
		Parameter & Value  \\
		\hline
		$H_0\,[\mathrm{km}\,\mathrm{s}^{-1}\,\mathrm{Mpc}^{-1}]$ & 67.27\\
		$\Omega_c$ & 0.2656 \\
		$\Omega_b$ & 0.04941 \\
		$\sigma_8$ & 0.8120 \\
        $n_s$ & 0.9649 \\
		\hline
	\end{tabular}
\end{table}
The factor $\bar{T}_{\rm HI} b_{\rm HI}$ enters the 2pCF as an overall normalization factor, which can be independently constrained. To isolate the impact of the instrumental beam on the clustering signal, we thus set $\bar{T}_{\rm HI} b_{\rm HI}=1$ throughout this analysis. 

The integration of equation \ref{eq:integral} 
involves highly oscillatory spherical Bessel functions and is thus numerically challenging. We perform this calculation using the {\tt FFTLog} algorithm as implemented in the Python package {\tt mcfit}\footnote{\url{https:/github.com/eelregit/mcfit}}, which is designed to provide stable and efficient Hankel-type transforms. We validate our implementation by comparing with the $\sigma=0\degr$ case computed using {\tt Hankl} \citep{Hankl:2021}, finding agreement at the level of better than 0.1 per cent over the range ($5-200\, {\rm  Mpc}\, h^{-1}$). For non-zero beam widths ($\sigma>0$), the beam smoothing further suppresses residual numerical oscillations, improving the stability of the calculation.

\begin{figure}
	\includegraphics[width=\linewidth]{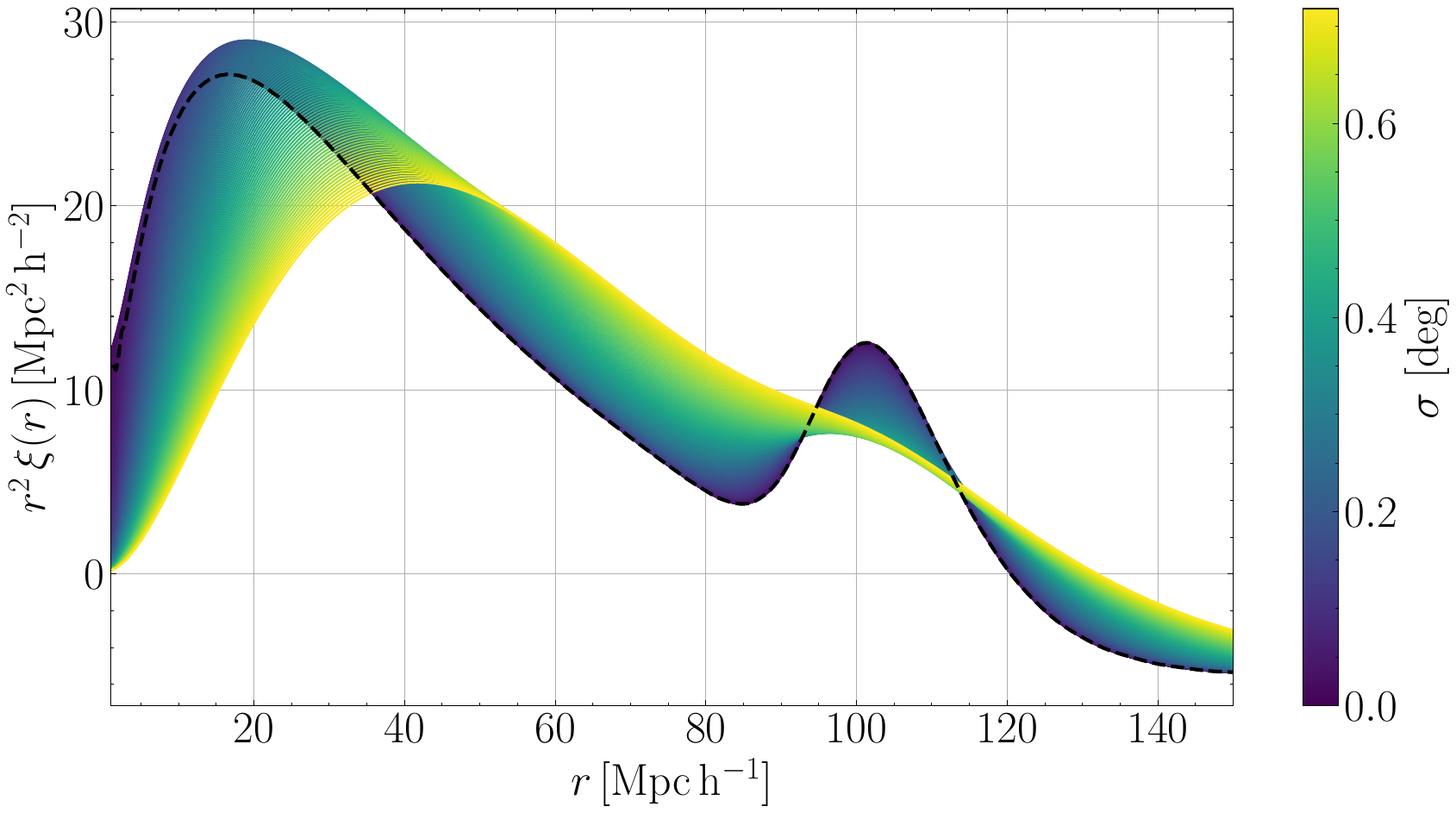}
   \caption{Illustration of the impact of different beam widths, $\sigma$, on the 2pCF at a fixed redshift, $z=0.8$.  The black dashed line corresponds to the $\sigma=0\degr$ case (no beam). All cases assume the fiducial $\Lambda$CDM cosmology.}
   \label{fig:xi}
\end{figure}

Figure~\ref{fig:xi} shows the 2pCF calculated for different beam widths, $\sigma$. Although the comoving size of the beam is not particularly large compared to $R_{\rm H}$ ($R_{\rm beam}=13.5\,\mathrm{Mpc}\,\mathrm{h}^{-1}$ for $\sigma=0.4\degr$), its impact extends across all scales. Two main effects can be identified in the 2pCF: (a) at small separations, the beam smooths the clustering signal, reducing the amplitude; the suppressed information is redistributed to larger scales, where the amplitude increases; and (b) the BAO peak becomes smeared at larger $\sigma$, illustrating the difficulty of recovering it from IM observations. Both conclusions are in agreement with results of \citet{2021/kennedy} and \citet{avila2022h}

\section{Transition scale to homogeneity}
\label{sec:homogeneity}

The homogeneity scale provides a critical observational test of the \bb{CP}. Although our measurement \bb{proves} only the past light cone rather than the entire Universe, the detection of this scale strongly supports one of the key assumptions of the $\Lambda$CDM model~\citep{2011/maartens}.

To quantify the transition to a homogeneous universe, one can adopt the scaled counts-in-spheres method. This approach measures the average number of tracers within a sphere of radius $r$, normalized by the expected count for a homogeneous distribution, defined as \citep{Scrimgeour:2012,Ntelis17}
\begin{equation}\label{eq:count_in_sphere}
\mathcal{N}(<r) \equiv \frac{N_{\rm gal}(<r)}{N_{\rm rand}(<r)},
\end{equation}
where $N_{\rm gal}(<r)$ denotes the mean counts within a sphere for a clustered distribution, while $N_{\rm rand}(<r)$ corresponds to the ideal homogeneous reference case.

Although the formalism was originally introduced for discrete tracers such as galaxies, it can be equivalently applied to continuous intensity fields, including 21\,cm IM~\citep{Camacho22}. At sufficiently large scales, $\mathcal{N}(<r)$ tends to unity, reflecting the onset of homogeneity. Since $N_{\rm gal}(<r)\propto r^{\mathcal{D}_{2}}$, where $D_2$ is the correlation dimension, and $N_{\rm rand}(<r)\propto r^{3}$, it follows that $\mathcal{N}(<r)\propto r^{\mathcal{D}_{2}-3}$. The correlation dimension can then be written as
\begin{equation}\label{eq:correlation_dimension}
    \mathcal{D}_2(r) \equiv \frac{d\ln\mathcal{N}(<r)}{d\ln r} + 3.
\end{equation}
At sufficiently large scales, $\mathcal{D}_2(r)$ approaches 3. Both $\mathcal{N}(<r)$ and $\mathcal{D}_2(r)$ can be used to estimate the homogeneity scale, however, \citet{Ntelis17} showed that $\mathcal{D}_{2}(r)$ is less correlated between scales than $\mathcal{N}(<r)$. As a result, the homogeneity scale inferred from $\mathcal{D}_2(r)$ is more robust.

To avoid numerical derivatives, which can introduce large errors, we instead express the scaled counts-in-spheres directly in terms of the 2pCF. 
In this formulation, $\mathcal{N}(<r)$ can be written as \citep{Ntelis17}:
\begin{equation}\label{eq:N_2pCF}
    \mathcal{N}(<r) \equiv 1 + \frac{3}{r^3}\int_{0}^{r}\xi(x)x^{2}dx = 1 + \bar{\xi}(r).
\end{equation}
where $\xi(r)$ is the monopole of the 2pCF and $\bar{\xi}(r)$ denotes its volume average. Substituting this expression in equation~(\ref{eq:correlation_dimension}), we obtain
\begin{equation}
    \mathcal{D}_2^{\rm obs}(\sigma,r) = 3\left[\frac{1+\xi^{\rm obs}(\sigma,r)}{1+\bar{\xi}^{\rm obs}(\sigma,r)}\right],
\end{equation}
where the dependence on the beam width, $\sigma$, has been explicitly included.

The transition to homogeneity is defined by the condition $\mathcal{D}_2^{\rm obs}(\sigma,R_{\rm H}^{\rm obs}) = 2.97$,
corresponding to a 1\% deviation from perfect homogeneity~\citep{Scrimgeour:2012}. Although somewhat arbitrary, this threshold has been widely adopted 
in the literature, enabling direct comparisons between analyses based on different tracers and methodologies. 

\begin{figure}
	\includegraphics[width=\linewidth]{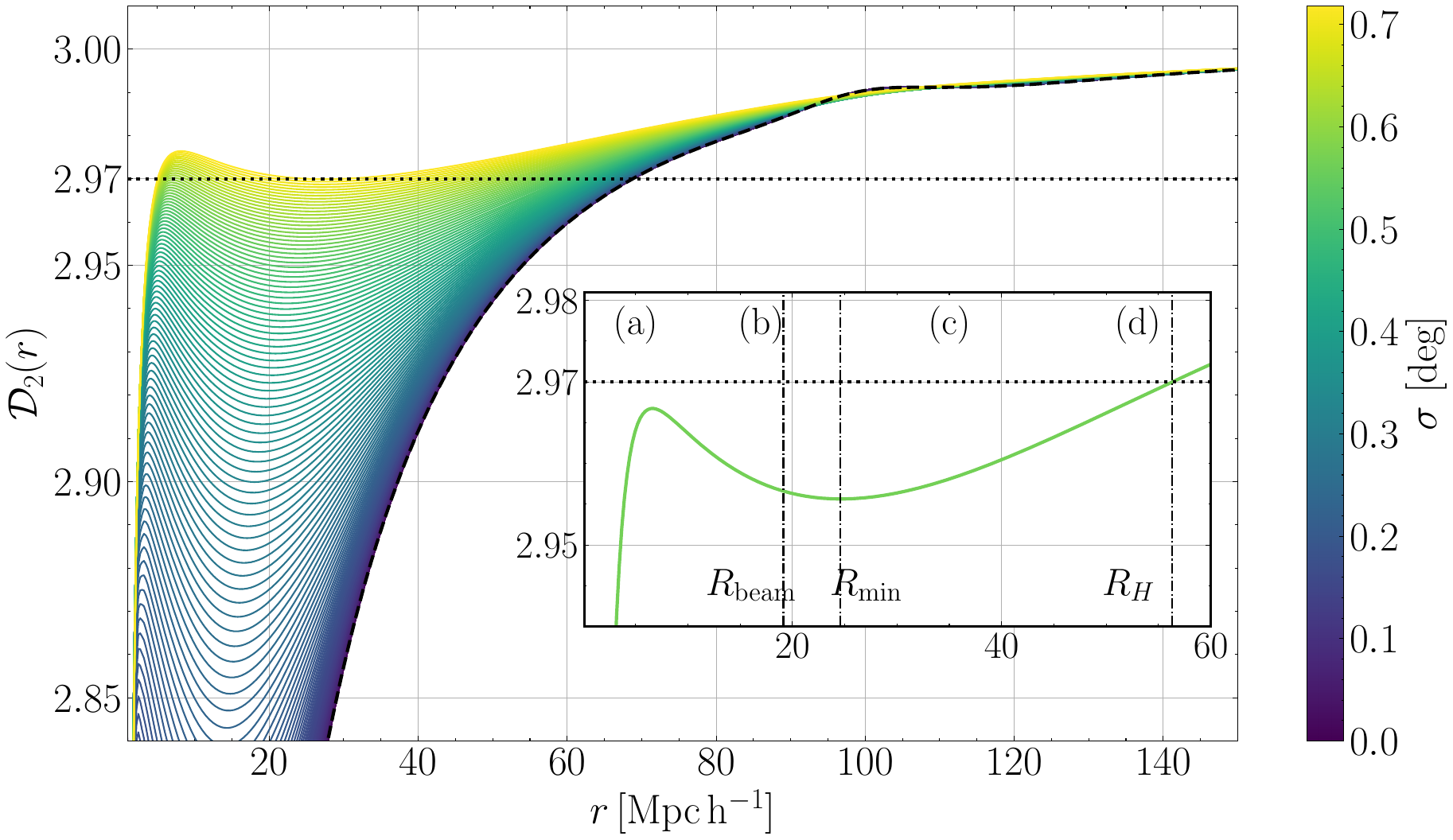}
   \caption{Effect of beam smoothing on the correlation dimension $\mathcal{D}_2(r)$. Curves correspond to increasing beam widths, with the black dashed line showing the $\sigma=0\degr$ case. All cases are calculated considering the fiducial $\Lambda$CDM cosmology, fixing $z=0.8$. The inner plot illustrates the four regimes (a–d) for $\sigma=0.5^\circ$, defined by the beam scale $R_{\rm beam}$, the scale of minimum $\mathcal{D}_2$, $R_{\rm min}$, and the homogeneity scale $R_{\rm H}$ where $\mathcal{D}_2$ crosses the threshold. See text for details.
   }
   \label{fig:d2_theory}
\end{figure}

Figure \ref{fig:d2_theory} presents the impact of the observational beam width, $\sigma$, on the correlation dimension, $\mathcal{D}_2$ illustrated for the redshift $z = 0.8$. The resulting curve is shaped by the interplay between the 21\,cm clustering, which lowers $\mathcal{D}_2$ at small scales, and beam-induced homogenization, which raises it. From this behaviour, we identify four distinct regimes, labelled as (a), (b), (c), and (d) in the inner plot of Figure \ref{fig:d2_theory}, delimited by $R_{\rm beam}$, $R_{\rm min}$ (the $r$ value at which $\mathcal{D}_2$ has a local minimum), and $R_{\rm H}$:\\

\paragraph*{(a) Beam-dominated Regime ($r\leq R_{\rm beam}$):} At the smallest scales, the measurement is dominated by the beam's smoothing kernel. Physical structures are washed out, leading to an artificially high value of $\mathcal{D}_2$ until it reaches a local maximum. As the effect of physical clustering emerges, $\mathcal{D}_2$ begins a downtrend.

\paragraph*{(b) Clustering dominated transition ($R_{\rm beam}\,<\,r\,\leq\,R_{\rm min}$):} As the scale $r$ exceeds the beam size ($R_{\rm beam}$), the beam's influence diminishes. This increased sensitivity to physical clustering causes $\mathcal{D}_2$ to decrease, reaching a minimum at $R_{\rm min}$.

\paragraph*{(c) Clustering-dominated Regime ($R_{\rm min}\,<\,r\,\leq\,R_{\rm H}$):} Beyond $R_{\rm min}$, $\mathcal{D}_2$ presents the usual behaviour, with a transition to homogeneity. In other words, as the distribution becomes progressively more uniform on larger scales, $\mathcal{D}_2$ rises toward the homogeneity scale, $R_H$.

\paragraph*{(d) Homogeneous Regime ($R_{\rm H}\,<\,r$):} 
At scales larger than that of the transition to homogeneity, $R_{\rm H}$, the 21\,cm would effectively be homogeneously distributed. This is reflected in the behaviour of the $\mathcal{D}_2$ curve, which asymptotically approaches the theoretical value of 3.

In summary, the primary impact of the instrumental beam on the 21\,cm signal clustering is an artificial homogenization (smoothing), directly affecting the transition scale. Even though the theoretical $R_{\rm H}$ is larger than the physical size of the beam, its value systematically decreases as the beam width increases. For sufficiently large beams, the correlation dimension can reach $\mathcal{D}_2 \simeq 2.97$ already at $R_{\rm min}$, effectively erasing any detectable transition to homogeneity in 21\,cm IM data. 
We note that, variations in the adopted threshold definition, $\mathcal{D}_2(\sigma,R_{\rm H}) = 2.97$, can lead to appreciable changes in the inferred values of $R_{\rm H}$, as ilustrated in the next section. However, since this threshold is conventional, its precise value does not alter the qualitative conclusions of our analysis.

These findings are discussed using the results obtained for the redshift $z = 0.8$ as an illustrative example. 
However, similar conclusions can be drawn from other redshift ranges, as we will discuss throughout the next section.

%%%%%%%%%%%%%%%%%%%%%%%%%%%%%%%%%%%%%%%%%%%%%%%%%%%%%%%%%%%%%
\section{\texorpdfstring{$R_{\rm H}$}{RH} detectability by future 21\,cm IM surveys}
%%%%%%%%%%%%%%%%%%%%%%%%%%%%%%%%%%%%%%%%%%%%%%%%%%%%%%%%%%%%%

\begin{figure}
	\includegraphics[width=\linewidth]{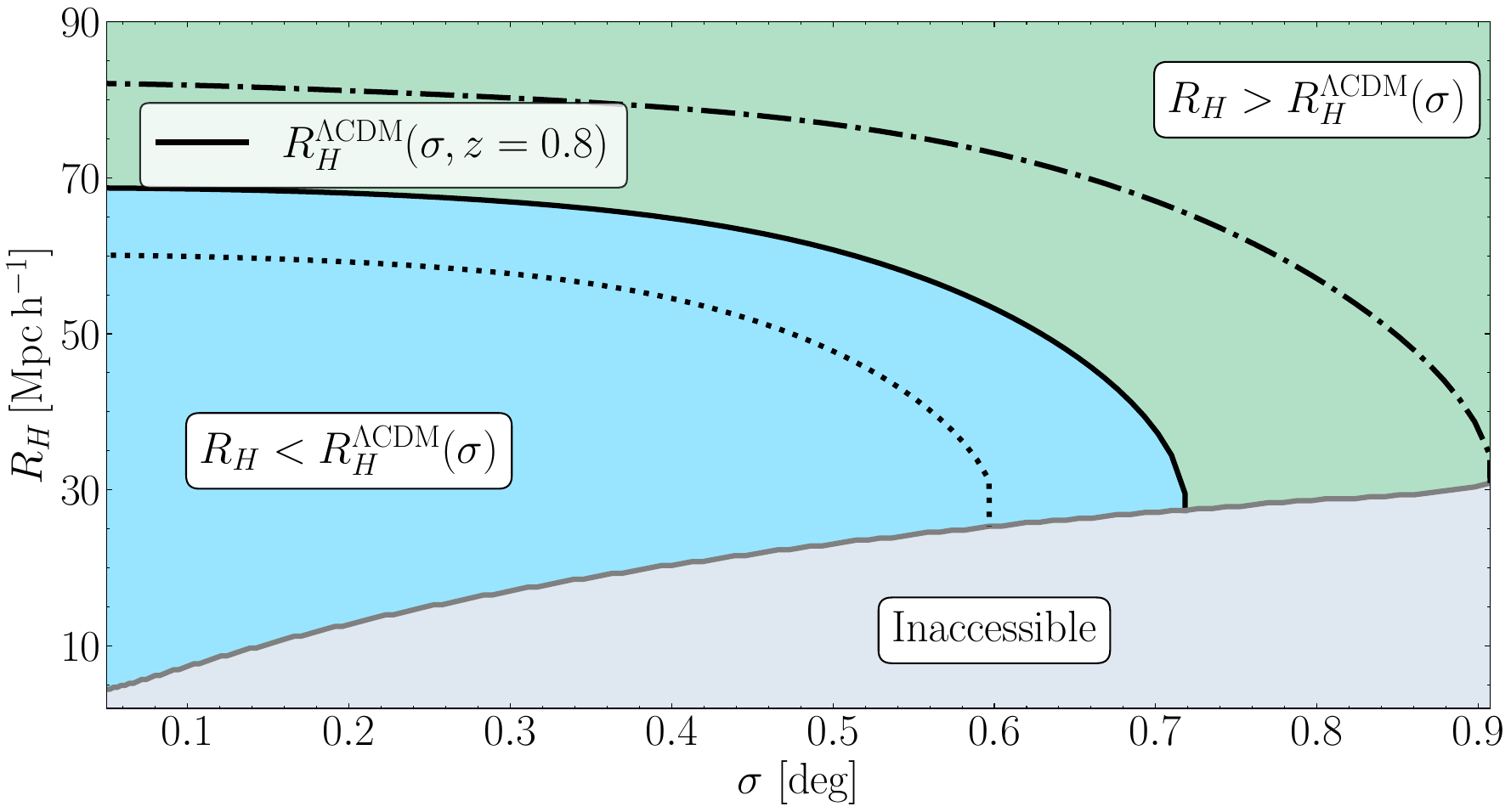}
   \caption{Detectability of the transition scale, $R_{\rm H}$, for varying beam widths ($\sigma$), illustrated for the fixed redshift, $z=0.8$.
   The black solid line represents $R^{\Lambda {\rm CDM}}_{\rm H}$ as affected by $\sigma$ considering the fiducial $\Lambda$CDM model. %\textcolor{red}{The black dotted and dash-dotted lines correspond to the homogeneity criteria of $\mathcal{D}_2(R_{\rm H})=2.96$ and $\mathcal{D}_2(R_{\rm H})=2.98$, respectively.}
   The black dotted and dash-dotted lines show the corresponding results obtained using the homogeneity criteria of $\mathcal{D}_2(R_{\rm H})=2.96$ and $\mathcal{D}_2(R_{\rm H})=2.98$, respectively.
   The grey solid line represents 
   %$R_{\rm H}(\sigma_{\rm max}) = R_{\rm min}(\sigma_{\rm max})$, with 
   $R^{\Lambda{\rm CDM}}_{\rm min}(\sigma)$}.
   %calculated for $\Lambda$CDM model. 
   The blue and green regions represent $R_{\rm H}$ values that could be measured, while the Inaccessible region, in grey, defines the $R_{\rm H}$ scales that would be erased by the beam effect. See text for details.
   \label{fig:homogeneity_window}
\end{figure}

In the previous section, we found that there is a maximum beam width, $\sigma=\sigma_{\rm max}$, up to which the homogeneity scale would still be detectable. This limiting width is given by the $\sigma$ value for which the homogeneity scale is reduced to the minimum scale, i.e., $R_{\rm H}(\sigma_{\rm max}) = R_{\rm min}(\sigma_{\rm max})$.  Figure \ref{fig:homogeneity_window} illustrates this limit as a solid dark-grey line, given by $R^{\Lambda{\rm CDM}}_{\rm min}(\sigma)$ calculated at different $\sigma$ values for the fiducial $\Lambda$CDM model at $z=0.8$. 
This curve defines the boundary of the so-called `Inaccessible' region (light grey area), where $\sigma$ is large enough to completely homogenize the 21\,cm signal.

The black solid line shows the theoretical prediction of $R_{\rm H}$ as a function of $\sigma$ within the fiducial $\Lambda$CDM cosmology. This curve also demonstrates the effect of the beam in lowering the $R_{\rm H}$ value; the wider the beam, the more significant this lowering. The blue and green regions represent the detectable $R_{\rm H}$ values for varying $\sigma$ widths.  %\textcolor{red}{Changes in the threshold value of 2.97 introduce considerable variations on the homogeneity scale as expected, but the values of $R_{\rm H}(\sigma)$ remains consistent with the inaccessible region, which does not depend on the threshold choice.}
Such results also show that, over these regions, the instrumental beam can be modelled to properly account for its effect on future measurements of the homogeneity scale using 21\,cm observations. 

For completeness, the black dotted and dot-dashed lines show the effect of adopting alternative homogeneity thresholds, $\mathcal{D}_2(R_{\rm H})$. 
While the $R_{\rm H}$ shifts accordingly, the boundary of the inaccessible region remains unchanged. 

Our finding about the existence of the threshold value $\sigma_{\rm max}$ for the beam width introduces a fundamental instrumental limitation on measurements of the homogeneity scale. By repeating our analysis over a range of redshift values, we can parametrize this maximum beam width as a function of redshift using the functional form
\begin{equation}\label{eq:sgima_max}
    \sigma_{\rm max}(z)\,=\,A + \frac{B}{z+C}.
\end{equation}
The best-fitting parameters (\textit{A}, \textit{B} and \textit{C}) are obtained using the Levenberg-Marquardt algorithm, implemented through the {\tt Scipy curve\_fit} routine \citep{2020/scipy}, applied to the discrete values of $\sigma_{\rm max}(z)$ calculated at each redshift. The fit is performed over the range $0.2 < z < 3$ with a redshift spacing of $\Delta z = 0.05$. All redshift points are weighted equally, assuming no correlation across them. The quoted uncertainties on the parameters are derived from the covariance matrix of the least-squares fit. The resulting best-fitting values are reported in Table \ref{tab:param}. The coefficient of determination, $R^2 \simeq 1$, confirms that this functional form provides an accurate description of the numerical results.
\begin{table}
	\centering
	\caption{Best-fitting parameters for equation \ref{eq:sgima_max}.}
	\label{tab:best}
	\begin{tabular}{lcccr} 
		\hline
		Type & A$\times 10^{2}$ & B$\times 10^{2}$ & C$\times 10^{2}$ & $R^2$\\
		\hline
		  RSD & -8.79$\pm$0.09 & 64.62$\pm$0.13 & 0.57$\pm$0.05 & $\approx 1$\\
		  no-RSD & -10.50$\pm$0.34 & 42.83$\pm$0.28 & -0.30$\pm$0.11 & $\approx 1$\\
		\hline
	\end{tabular}
\end{table}

The quantity $\sigma_{\rm max}(z)$ corresponds to the beam width at which the transition to homogeneity becomes indistinguishable from the minimum of the correlation dimension, i.e. where $R_{\rm H} = R_{\rm min}$. This condition is illustrated in Figure~\ref{fig:homogeneity_window} by the intersection between the $R^{\Lambda{\rm CDM}}_{\rm H}$ and $R^{\Lambda{\rm CDM}}_{\rm min}$ curves at the redshift $z=0.8$.

\begin{figure*}
	\includegraphics[width=0.8\linewidth]{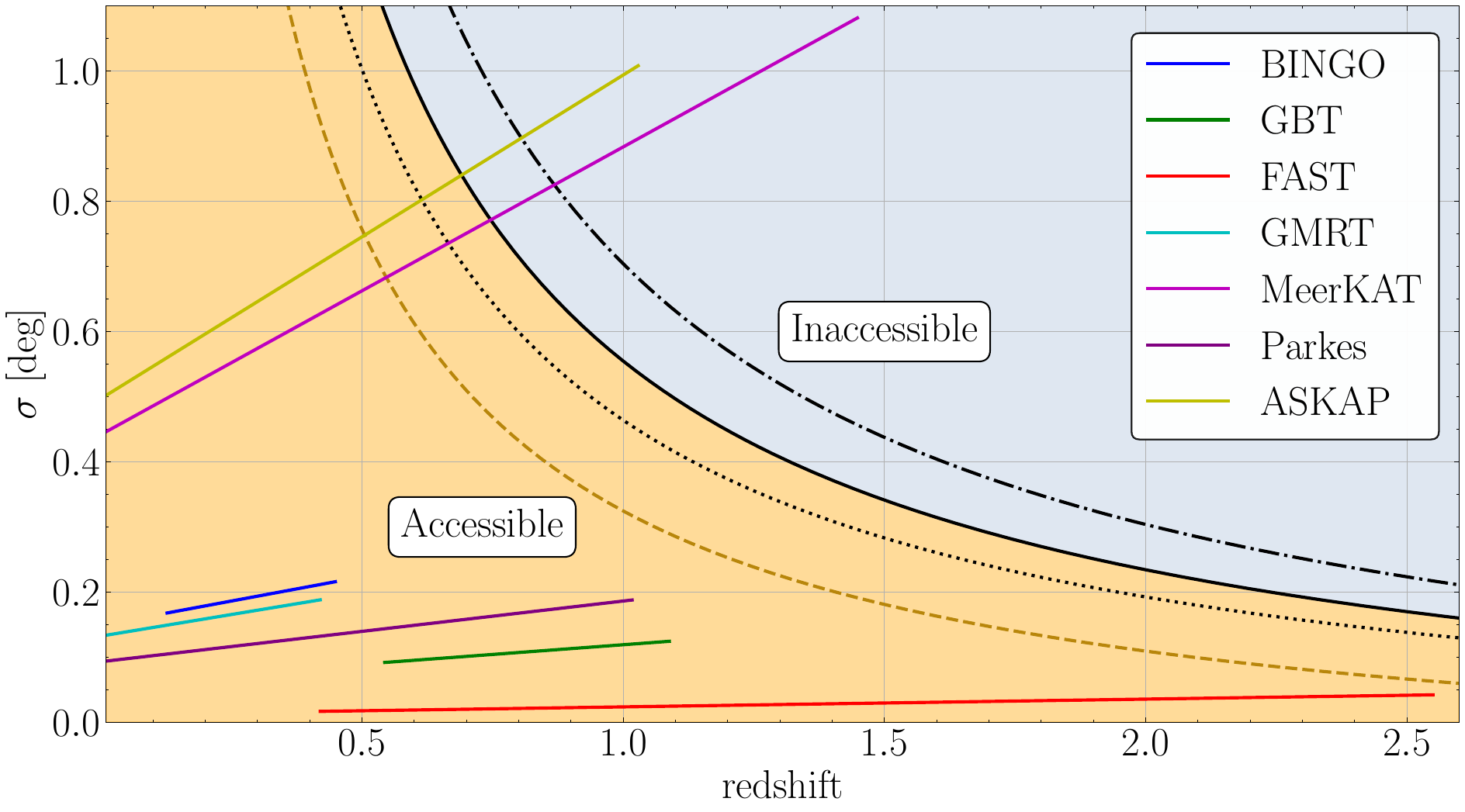}
   \caption{Detectability of $R_{\rm H}$ 
   in the $\sigma - z$ parameter space.
   The black solid and gold dashed lines represent $\sigma_{\rm max}(z)$, as fitted by equation \ref{eq:sgima_max} for the $\Lambda$CDM model when the Kaiser effect is included and neglected, respectively. The black dotted and dash-dotted lines 
   show results (including Kaiser effect) for
   the homogeneity criteria of $\mathcal{D}_2(R_{\rm H})=2.96$ and $\mathcal{D}_2(R_{\rm H})=2.98$, respectively. The coloured lines show the $\sigma(z)$ for several 21\,cm IM instruments. Their redshift (frequency) ranges are summarised in Table \ref{tab:instruments} \bb{.} $\sigma(z)$ values \bb{were} calculated using equation \ref{eq:beam_width}.}
   \label{fig:instruments}
\end{figure*}

This framework allows us to determine whether a given instrument can access the homogeneity scale over its operational frequency band, given its beam widths. 
Figure \ref{fig:instruments} presents the $\sigma - z$ parameter space, showing $\sigma_{\rm max}(z)$ curves. The curve delimits the orange and grey regions, which correspond, respectively, to $\sigma > \sigma^{\rm \tiny \Lambda CDM}_{\rm max}$, inaccessible to $R_{\rm H}$ measurements, and to $\sigma < \sigma^{\rm \tiny \Lambda CDM}_{\rm max}$, where $R_{\rm H}$ would be detectable. In addition, we display the $\sigma_{\rm max}(z)$ curve for the case without RSD, where it returns lower values. This occurs because the RSD causes the clustering to appear more intense, increasing $\sigma_{\rm max}(z)$. 
The corresponding best-fitting parameters, A, B, and C, for the no-RSD case are also reported in Table \ref{tab:best}. 
We compare these limits with the expected beam performance for several single-dish radio telescopes (see Table \ref{tab:instruments} for specifications).
Finally, we ilustrate the impact of adopting alternative homogeneity thresholds, $\mathcal{D}_2(R_{\rm H})$. 
Although different choices shift the numerical value of $R_{\rm H}$ and, consequently, $\sigma_{\rm max}$, our qualitative conclusions regarding detectability of $R_{\rm H}$ and the existence of an inaccessible region remain unaffected.

Instruments operating in interferometric mode were not considered in this work, as their synthesized beams are typically much narrower than those of single-dish experiments. The angular resolution of an interferometer is approximately given by $\sigma_{\rm synth} \simeq \lambda/B_{\rm max}$, where $B_{\rm max}$ is the maximum baseline length. For current and upcoming 21\,cm IM interferometers, $B_{\rm max}$ ranges from hundreds of meters 
 \cite[e.g. HERA, HIRAX, and CHIME; see][respectively]{deboer2017hydrogen,newburgh2016hirax,2014/chime}
%[e.g. HERA \citep{deboer2017hydrogen}, HIRAX \citep{newburgh2016hirax}, CHIME \citep{2014/chime}]
%to several kilometres 
%[SKA-low and SKA-mid \citep{2020/ska}]. 
% \cite[SKA-low and SKA-mid; ][]{2020/ska}. 
As a result, interferometric observations lie firmly within the 'Accessible' region of the $\sigma - z$ space, since their synthesized beam widths satisfy $\sigma_{\rm synth} \ll \sigma_{\rm max}(z)$ over the relevant redshift range. This reinforces the practical relevance of our framework: while single-dish experiments must carefully account for beam-induced homogenization effects, interferometers are naturally well positioned to recover the transition to homogeneity with high precision, provided that foreground contamination and thermal noise are sufficiently controlled.
\begin{table}
    \centering
    \caption{\bb{Specifications for the single-dish and interferometric instruments considered in this work.}} 
    \label{tab:instruments}
    \begin{tabular}{lccc} 
        \hline
        \multirow{2}{25pt}{Instrument} & \multirow{2}{40pt}{\centering $\nu_{\rm min} - \nu_{\rm max}$ [MHz]} & \multirow{2}{45pt}{\centering Dish diameter [m]} & \multirow{2}{25pt}{Ref.} \\
         & & & \\
        \hline
        BINGO   & 980--1260 & 40 & \citet{2021/wuensche_BINGO-instrument}\\
        GBT     & 680--920  & 100 & \citet{Masui:2013}\\
        FAST    & 400--1000 & 500 & \citet{2011/fast}\\
        GMRT    & 1000--1500 & 45 & \citet{patra2019expanded}\\
        MeerKAT & 580--1680 & 13.5 & \citet{santos2017meerklass}\\
        Parkes  & 704--4032 & 64 & \citet{hobbs2020ultra}\\
        ASKAP   & 700--1800 & 12 & \citet{ASKAP2008}\\
        \hline 
    \end{tabular}
\end{table}
%\footnotetext[3]{\citet{2021/wuensche_BINGO-instrument}}
%\footnotetext[4]{\citet{Masui:2013}}
%\footnotetext[5]{\citet{2011/fast}}
%\footnotetext[6]{\citet{patra2019expanded}}
%\footnotetext[7]{\citet{santos2017meerklass}}
%\footnotetext[8]{\citet{hobbs2020ultra}}
%\footnotetext[9]{\citet{ASKAP2008}}

%

Among the single-dish instruments investigated, the ability of ASKAP and MeerKAT to measure $R_{\rm H}$ is clearly impacted. Due to their relatively small dish diameter (12 m and 13.5 m, respectively), their beams become too wide at high redshifts (low frequencies), making them unable to probe the homogeneity scale at $z \gtrsim 0.6$, though measurements at lower redshifts remain possible. All other instruments appear fully inside the accessible region and would have the potential to measure the transition scale to homogeneity to provide consistency tests of the CP.

Finally, it is important to note that the Gaussian beam model employed here should be regarded as a first-order approximation. 
A realistic beam profile, accounting for features such as sidelobes, would likely produce a stronger homogenizing effect on the observed signal. 
As a result, a more realistic inaccessible region is expected to be larger, extending to smaller values of $\sigma$ than those predicted by our idealized model.

%%%%%%%%%%%%%%%%%%%%%%%%%%%%%%%%%%
\section{Conclusions}
%%%%%%%%%%%%%%%%%%%%%%%%%%%%%%%%%%

This work presents the first assessment of the feasibility of measuring the transition scale to cosmic homogeneity, $R_{\rm H}$, with current and future 21\,cm IM surveys. In particular, we investigate how the instrumental beam affects this measurement. We provide a detailed interpretation of how different beam widths affect the two-point correlation function, artificially homogenizing smaller scales, and discuss the resulting impact on the correlation dimension, $D_{2}(r)$, used to estimate $R_{\rm H}$.

We show that such a measurement is, in principle, achievable, though constrained by instrumental limitations determined by the telescope beam width and observing frequency. Our main contribution is a theoretical framework that quantifies these limitations and defines a criterion for the maximum beam width, $\sigma_{\rm max}(z)$, \bb{given by equation~\ref{eq:sgima_max}}, below which $R_{\rm H}$ remains accessible at a given redshift. This limit naturally separates the accessible and inaccessible regions in the $\sigma - z$ parameter space, where the homogeneity signal can be recovered or is erased by beam smoothing. We further assess the capability of several instruments to perform such measurements.

This study lays the theoretical foundation for more comprehensive future analyses. 
Next steps include applying our framework to realistic simulations and, ultimately, to data from forthcoming surveys. While we have focused on the beam's impact, future work should also address other observational challenges, such as foreground contamination and instrumental systematics \citep{Matshawule:2021, Amiri_chime:2024, cunnington_meerklass:2025}. In this context, cross-correlating 21\,cm IM data with overlapping galaxy surveys may prove particularly valuable, as it can enhance the signal-to-noise ratio and mitigate systematics similarly to its role in power-spectrum analyses. Finally, we emphasize that the framework presented here is not limited to the 21\,cm line and can be readily adapted to study cosmic homogeneity with other line IM tracers.

\section*{Acknowledgements}
The authors acknowledge Phil Bull for insightful discussions. 
We also thank an anonymous referee for relevant comments.
B.B.B. acknowledges resources received from Fundação de Amparo à Pesquisa do Estado de São Paulo (FAPESP) under processes 2022/16749-0 and 2024/12902-3. 
C.P.N. thanks Serrapilheira Institute. 
F.A thanks Fundação Carlos Chagas Filho de Amparo à Pesquisa do Estado do Rio de Janeiro (FAPERJ), Process SEI-260003/001221/2025, for the financial support. 
R.M. acknowledges the financial support from Conselho Nacional de Desenvolvimento Científico e Tecnológico (CNPq) under the fellowship 302370/2024-2. 
H.H.C thanks Coordenação de Aperfeiçoamento de Pessoal de Nível Superior (CAPES) for grant 88887.212226/2025-00 under DS program.
C.A.W. thanks CNPq for grants 312505/2022-1 and 407446/2021-4. 
G.A.S.S. acknowledges resources received from CAPES under process 88887.602823/2021-00.
The results of this work used the software packages {\tt numpy} \citep{2011/numpy}, {\tt scipy} \citep{2020/scipy} and {\tt matplotlib} \citep{2007/matplotlib}.

%%%%%%%%%%%%%%%%%%%%%%%%%%%%%%%%%%%%%%%%%%%%%%%%%%
\section*{Data Availability}

The code developed to calculate the 2pCF's and produce the data underlying this article are publicly available at \url{github.com/Bruno-Bizarria/2pIM}.



\bibliographystyle{mnras}
%\bibliography{example}
\bibliography{hom_scale_ref}% if your bibtex file is called example.bib

% Don't change these lines
\bsp	% typesetting comment
\label{lastpage}
\end{document}